\documentclass[journal]{IEEEtran}
\IEEEoverridecommandlockouts

\usepackage{xspace}
\usepackage{glossaries}
\usepackage{color}

\usepackage{graphicx}        
\graphicspath{{./figures/}}
\usepackage{subfigure}

 \usepackage{caption}
\usepackage{cite}

\let\OLDthebibliography\thebibliography
\renewcommand\thebibliography[1]{
	\OLDthebibliography{#1}
	\setlength{\parskip}{0pt}
	\setlength{\itemsep}{0pt}
}

\usepackage[detect-weight=true, binary-units=true]{siunitx}
\usepackage{amsmath,amssymb,latexsym}
\usepackage{amsfonts}
\usepackage{upgreek}
\usepackage{mathtools}
\usepackage{nicefrac}

\usepackage{adjustbox}
\usepackage{multirow}
\usepackage{array}
   \newcolumntype{P}[1]{>{\centering\arraybackslash}p{#1}}

\usepackage{algorithm}
\usepackage{algorithmicx}
\usepackage{algcompatible}
\usepackage[noend]{algpseudocode}
\renewcommand{\eqref}[1]{(\ref{#1})}
\newcommand{\secref}[1]{\mbox{Section~\ref{#1}}}

\newcommand{\figref}[1]{\mbox{Fig.~\ref{#1}}}
\newcommand{\tblref}[1]{\mbox{Table~\ref{#1}}}

\newcommand{\blue}[1]{\textcolor{black}{#1}}
\newcommand{\black}[1]{\textcolor{black}{#1}}

\newacronym{enics}{EnICS}{Emerging NanoScaled Circuits \& Systems}

\newacronym{biu}{BIU}{Bar-Ilan University}

\newacronym{snr}{SNR}{signal-to-noise ratio}
\newacronym{mep}{MEP}{minimum energy point}
\newacronym{ppa}{PPA}{performance, power, and area}

\newacronym{mu}{$\mu$}{mean}
\newacronym{sigma}{$\sigma$}{standard deviation}
\newacronym{ser}{SER}{soft errors}
\newacronym{seu}{SEU}{single-event upset}
\newacronym{cdf}{CDF}{cumulative distribution function}
\newacronym{pdf}{PDF}{probabililty distribution function}

\newacronym{itrs}{ITRS}{International Technology Roadmap for Semiconductors}
\newacronym{ic}{IC}{integrated circuit}
\newacronym{ip}{IP}{intellectual property}
\newacronym{soi}{SOI}{silicon-on-insulator}
\newacronym{fdsoi}{FD-SOI}{fully-depleted silicon-on-insulator}

\newacronym{sc}{SC}{standard cell}
\newacronym{soc}{SoC}{system-on-chip}
\newacronym{vdd}{$V_{\text{DD}}$}{supply voltage}
\newacronym{vboost}{$V_{\text{BOOST}}$}{boosted voltage}
\newcommand{\vdd}{$V_{\text{DD}}$\xspace}

\newacronym{gnd}{$GND$}{ground}
\newacronym{vtc}{VTC}{voltage transfer characteristic}
\newacronym{rdf}{RDF}{random dopant fluctuations}
\newacronym{ulp}{ULP}{ultra-low power}
\newacronym{pvt}{PVT}{Process-Voltage-Temperature}
\newacronym{vlsi}{VLSI}{very large scale integration}
\newacronym{asic}{ASIC}{application specific integrated circuit}
\newacronym{pcb}{PCB}{printed circuit board}
\newacronym{fir}{FIR}{finite impulse response}
\newacronym{dsp}{DSP}{digital signal processing}
\newacronym{lut}{LUT}{look-up table}
\newacronym{mac}{MAC}{multiply-and-accumulate}

\newacronym{cmos}{CMOS}{complementary-metal-oxide-semiconductor}
\newacronym{dff}{DFF}{Data Flip-Flop}

\newacronym{subvt}{sub-$V_{\text{T}}$}{sub-threshold}

\newacronym{nearvt}{near-$V_{\text{T}}$}{near threshold}
\newacronym{vt}{$V_{\text{T}}$}{threshold voltage}

\newacronym{vgs}{$V_{\text{GS}}$}{gate-to-source voltage} 
\newacronym{vds}{$V_{\text{DS}}$}{drain-to-source voltage} 
\newacronym{vbs}{$V_{\text{BS}}$}{body-to-source voltage} 
\newacronym{vgb}{$V_{\text{GB}}$}{gate-to-body voltage} 
\newacronym{dibl}{DIBL}{drain induced barrier lowering}
\newacronym{gidl}{GIDL}{gate induced drain leakage} 
\newacronym{ids}{$I_{\text{DS}}$}{drain-to-source current}
\newacronym{sce}{SCE}{short channel effect}
\newacronym{rsce}{RSCE}{reverse short channel effect}
\newacronym{tox}{$t_{\text{ox}}$}{gate oxide thickness}
\newacronym{L}{$L$}{channel length}
\newacronym{W}{$W$}{channel width}
\newacronym{rbb}{RBB}{reverse body biasing}
\newacronym{fbb}{FBB}{forward body biasing}
\newacronym{btbt}{BTBT}{band-to-band tunneling}
\newacronym{bjt}{BJT}{bipolar junction transistor}
\newacronym{hvt}{HVT}{high threshold voltage}
\newacronym{lvt}{LVT}{low threshold voltage}
\newacronym{nvt}{NVT}{nominal threshold voltage}
\newacronym{pmos}{PMOS}{p-type MOSFET}
\newacronym{nmos}{NMOS}{n-type MOSFET}
\newacronym{isub}{$I_\text{sub}$}{sub-threshold leakage}
\newacronym{igate}{$I_\text{gate}$}{gate leakage}
\newacronym{ibulk}{$I_\text{bulk}$}{bulk leakage}
\newacronym{vbb}{$V_{\text{BB}}$}{body voltage} 

\newacronym{2T}{2T}{two-transistor}
\newacronym{3T}{3T}{three-transistor}
\newacronym{4T}{4T}{four-transistor}
\newacronym{6T}{6T}{six-transistor}
\newacronym[longplural={static random access memories}]{sram}{SRAM}{static random access memory}

\newacronym{1T1C}{1T-1C}{1-transistor 1-capacitor}
\newacronym[longplural={dynmaic random access memories}]{dram}{DRAM}{dynamic random access memory}

\newacronym{ssd}{SSD}{solid-state hard drive}
\newacronym{L1}{L1}{level-1}
\newacronym{L2}{L2}{level-2}
\newacronym{L3}{L3}{level-3}
\newacronym{L4}{L4}{level-4}
\newacronym{nvm}{NVM}{non-volatile memory}
\newacronym{rom}{ROM}{read-only memory}
\newacronym{fifo}{FIFO}{first-in first-out}
\newacronym{lifo}{LIFO}{last-in first-out}

\newcommand{\one}{`1'\xspace}

\newacronym{wl}{WL}{word line}
\newcommand{\wl}{\gls{wl}\xspace}
\newacronym{wbl}{WBL}{write bit line}

\newacronym{rbl}{RBL}{read bit line}

\newacronym{mw}{MW}{write transistor}
\newacronym{sn}{SN}{storage node}

\newacronym{wwl}{WWL}{write word line}

\newacronym{rwl}{RWL}{read word line}

\newacronym{mux}{mux}{multiplexer}
\newacronym{mlc}{MLC}{multi-level cell}
\newacronym{drv}{DRV}{data retention voltage}
\newacronym{nwl}{NWL}{negative word line}
\newacronym{bist}{BIST}{built-in self-test}
\newacronym{bisr}{BISR}{built-in self-repair}
\newacronym{reram}{Re-RAM}{resistive RAM}
\newacronym{ecc}{ECC}{error correction code}
\newacronym{cdmr}{CDMR}{complementary dual-modular redundancy}

\newacronym{snm}{SNM}{static noise margin}
\newacronym{rsnm}{RSNM}{read static noise margin}
\newacronym{wsnm}{WSNM}{write static noise margin}
\newacronym{dnm}{DNM}{dynamic noise margin}

\newacronym{qcrit}{$Q_\text{crit}$}{critical charge}
\newacronym{dmr}{DMR}{dual modular redundancy}
\newacronym{edac}{EDAC}{error detection and correction}
\newacronym{secded}{SECDED}{single error correction~-- double error detection}
\newacronym{dected}{DECTED}{double error correction~-- triple error detection}

\newacronym{gc}{GC}{gain cell}
\newacronym{edram}{eDRAM}{embedded DRAM}
\newacronym{gcedram}{GC-eDRAM}{gain-cell embedded DRAM}

\newacronym{drt}{DRT}{data retention time}

\newacronym{dcvsl}{DCVSL}{differential cascade voltage switch logic}

\newacronym{dif}{DIF}{digital implementation flow}
\newacronym{hdl}{HDL}{hardware description language}
\newacronym{rtl}{RTL}{register transfer level}
\newacronym{eda}{EDA}{electronic design automation}
\newacronym{cad}{CAD}{computer-aided design}
\newacronym{pr}{P\&R}{place and route}
\newacronym{cts}{CTS}{clock-tree synthesis}
\newacronym{sta}{STA}{static timing analysis}
\newacronym{edi}{EDI}{Cadence Encounter Design Implementation}
\newacronym{s_dc}{DC}{Synopsys Design Compiler}
\newacronym{sdc}{SDC}{Synopsys Design Constraints}
\newacronym{vcd}{VCD}{value change dump}

\newacronym{scm}{SCM}{standard cell memory}

\newacronym{smu}{SMU}{source/measure unit}
\newacronym{dmm}{DMM}{digital multimeter}

\newacronym{lpa}{LPA}{Leakage Power Analysis}
\newacronym{dpa}{DPA}{Differential Power Analysis}

\newacronym{puf}{PUF}{Physical Unclonable Function}

\newacronym{dl}{DL}{deep learning}

\newacronym{ai}{AI}{artificial intelligence}

\newacronym{cnn}{CNN}{convolutional neural network}

\newacronym{dnn}{DNN}{deep neural network}

\newacronym{cpu}{CPU}{central processing unit}

\newacronym{gpu}{GPU}{graphics processing unit}

\newacronym{tpu}{TPU}{tensor processing unit}

\newacronym{relu}{ReLu}{rectified linear unit}
\usepackage[hidelinks]{hyperref}
\usepackage{caption}
\usepackage[font=small]{caption} 
\usepackage{float} 

\newcommand{\slbar}{$\overline{\text{SL}}$\xspace} 
\newcommand{\PCML}{$\overline{\text{PC-ML}}$\xspace} 
\newcommand{\PCMLnospace}{$\overline{\text{PC-ML}}$\xspace} 
\newcommand{\McOne}{$M_{\text{C1}}$\xspace} 
\newcommand{\McTwo}{$M_{\text{C2}}$\xspace} 
\newcommand{\McThree}{$M_{\text{C3}}$\xspace} 
\newcommand{\data}{$\text{D}$\xspace} 
 
\newcommand{\Meval}{$M_{\text{eval}}$\xspace}
\newcommand{\Veval}{$V_{\text{eval}}$\xspace}
\newcommand{\Vevalth}{$V_{\text{evalth}}$\xspace}
\newcommand{\teval}{$t_{\text{eval}}$\xspace}

\newacronym{tt}{TT}{Typical-Typical}
    \renewcommand{\tt}{\gls{tt}\xspace} 
\newacronym{ss}{SS}{Slow-Slow}
    \renewcommand{\ss}{\gls{ss}\xspace} 
\newacronym{ff}{FF}{Fast-Fast}
    \newcommand{\ff}{\gls{ff}\xspace}     
\newacronym{mc}{MC}{Monte Carlo}
    \newcommand{\mc}{\gls{mc}\xspace}  
\newacronym{hd}{HD}{Hamming distance}
    \newcommand{\hd}{{Hamming distance}\xspace}

\newacronym{ngs}{NGS}{next-generation sequencing}

\newacronym{ncbi}{NCBI}{National Center for Biotechnology Information}
    \newcommand{\ncbi}{\gls{ncbi}\xspace} 
\newacronym{covid}{SARS-CoV-2}{Severe Acute Respiratory Syndrome coronavirus-2}
    \newcommand{\covid}{\gls{covid}\xspace}    
\newacronym{pcr}{PCR}{polymerase chain reaction}
    \newcommand{\pcr}{\gls{pcr}\xspace}

\newacronym[longplural={six-transistor static random access memories}]{sixtsram}{6T-SRAM}{six-transistor static random access memory}
    \newcommand{\sixtsram}{\gls{sixtsram}\xspace}

\newacronym[longplural={magnetic random-access memories}]{mram}{MRAM}{magnetic random-access memory}

\newacronym[longplural={content-addressable memories}]{cam}{CAM}{content-addressable memory}

\newacronym[longplural={spin-transfer torque magnetic random-access memories}]{sttmram}{STT-MRAM}{spin-transfer torque magnetic random-access memory}

\newacronym{bc}{BC}{bitcell}

\newacronym{ml}{ML}{matchline}
    \newcommand{\ml}{\gls{ml}\xspace}

\newacronym{sa}{SA}{sense amplifier}    
    \newcommand{\sa}{\gls{sa}\xspace}

\newacronym{sl}{SL}{searchline}    
    \renewcommand{\sl}{\gls{sl}\xspace} 
     
    \newcommand{\sls}{\glspl{sl}\xspace}
    
\newacronym{pc}{PC}{pre-charge}    
    \newcommand{\pc}{\gls{pc}\xspace}

\newacronym{pcml}{\PCML}{\ml pre-charge}    
    \newcommand{\pcml}{\gls{pcml}\xspace} 
\newacronym{pcsl}{PC-SL}{\sl pre-charge}    
    \newcommand{\pcsl}{\gls{pcsl}\xspace}   
\newacronym{mlsa}{MLSA}{matchline sense-amplifier}    
    \newcommand{\mlsa}{\gls{mlsa}\xspace} 
    
\newacronym{bl}{BL}{bit line}

\newacronym{sot}{SOT}{spin-orbit torque}

\newacronym{she}{SHE}{spin Hall effect}

\newacronym{stt}{STT}{spin-transfer torque}

\newacronym{vcma}{VCMA}{voltage-controlled magnetic anisotropy}

\newacronym{mtj}{MTJ}{magnetic tunnel junction}

\newacronym{smtj}{SMTJ}{single-barrier magnetic tunnel junction}

\newacronym{dmtj}{DMTJ}{double-barrier MTJ}

\newacronym{amr}{AMR}{anisotropic magnetoresistance}

\newacronym{gmr}{GMR}{giant magnetoresistance}

\newacronym{tmr}{TMR}{tunnel magnetoresistance}

\newacronym{mr}{MR}{magnetoresistance}

\newacronym{wer}{WER}{write error rate}

\newacronym{fl}{FL}{free layer}

\newacronym[longplural={reference layers}]{rl}{RL}{reference layer}

\newacronym{p}{P}{parallel}
    
\newacronym{ap}{AP}{antiparallel}

\newacronym{fm}{FM}{ferromagnetic}

\newcommand{\kmers}{$k$-mers\xspace}
\newcommand{\kmer}{$k$-mer\xspace}

\newacronym{ASIen}{ASI}{Italian Space Agency}

\newacronym{ASIit}{ASI}{Agenzia Spaziale Italiana}

\newacronym{UNICALen}{UNICAL}{University of Calabria}

\newacronym{UNICALit}{UNICAL}{Università della Calabria}

\newacronym{DIMESen}{DIMES}{Department of Computer Engineering, Modeling, Electronics and Systems}

\newacronym{DIMESit}{DIMES}{Dipartimento di Ingegneria Informatica, Modellistica, Elettronica e Sistemistica}

\newcommand{\V}{\,\si{\volt}\xspace}

\newcommand{\nm}{\,\si{\nano\meter}\xspace}

\newcommand{\ps}{\,\si{\pico\second}\xspace}
\newcommand{\ns}{\,\si{\nano\second}\xspace}

\newcommand{\bit}{\,\si{\bit}\xspace}

\newcommand{\Xnospace}{$\times$}
\usepackage{scalerel}
\usepackage{tikz}
\usetikzlibrary{svg.path}
\definecolor{orcidlogocol}{HTML}{A6CE39}
\tikzset{
  orcidlogo/.pic={
    \fill[orcidlogocol] svg{M256,128c0,70.7-57.3,128-128,128C57.3,256,0,198.7,0,128C0,57.3,57.3,0,128,0C198.7,0,256,57.3,256,128z};
    \fill[white] svg{M86.3,186.2H70.9V79.1h15.4v48.4V186.2z}
                 svg{M108.9,79.1h41.6c39.6,0,57,28.3,57,53.6c0,27.5-21.5,53.6-56.8,53.6h-41.8V79.1z M124.3,172.4h24.5c34.9,0,42.9-26.5,42.9-39.7c0-21.5-13.7-39.7-43.7-39.7h-23.7V172.4z}
                 svg{M88.7,56.8c0,5.5-4.5,10.1-10.1,10.1c-5.6,0-10.1-4.6-10.1-10.1c0-5.6,4.5-10.1,10.1-10.1C84.2,46.7,88.7,51.3,88.7,56.8z};
  }
}

\newcommand\orcidicon[1]{\href{https://orcid.org/#1}{\mbox{\scalerel*{
\begin{tikzpicture}[yscale=-1,transform shape]
\pic{orcidlogo};
\end{tikzpicture}
}{|}}}}

\newcommand{\GSMieee}{\IEEEmembership{Graduate~Student~Member,~IEEE}}
\newcommand{\Mieee}{\IEEEmembership{Member,~IEEE}}
\newcommand{\SMieee}{\IEEEmembership{Senior~Member,~IEEE}}

\newcommand{\UNICALfull}{Department of Computer Engineering, Modeling, Electronics and Systems, University of Calabria, Rende 87036, Italy\xspace}

\newcommand{\BIUwEnICSfull}{Emerging Nanoscaled Integrated Circuits \& Systems (EnICS) Labs, Faculty of Engineering, Bar-Ilan University, Ramat-Gan 5290002, Israel\xspace}

\newcommand{\Technionfull}{Department of Electrical Engineering, Technion-Israel Institute of Technology, Haifa 3547902, Israel\xspace}

\newcommand{\EGnospace}{\,Esteban~Garzón}
\newcommand{\EGorcid}{\,\orcidicon{0000-0002-5862-2246}}

\newcommand{\ATnospace}{\,Adam~Teman}
\newcommand{\ATorcid}{\,\orcidicon{0000-0002-8233-4711}}

\newcommand{\MLnospace}{\,Marco~Lanuzza}
\newcommand{\MLorcid}{\,\orcidicon{0000-0002-6480-9218}}

\newcommand{\RGnospace}{\,Roman~Golman}
\newcommand{\RGorcid}{\,\orcidicon{0000-0002-1215-5603}}

\newcommand{\LYnospace}{\,Leonid~Yavits}
\newcommand{\LYorcid}{\,\orcidicon{0000-0001-5248-3997}}

\newcommand{\NVnospace}{\,Natan~Vinshtok-Melnik}
\newcommand{\NVorcid}{\,\orcidicon{0000-0002-8848-9941}}

\usepackage{subfiles} 

\begin{document}

\title{\black{Hamming Distance Tolerant Content-Addressable Memory (HD-CAM) for Approximate Matching Applications}}

\author{
        \EGnospace$^{\EGorcid}$, \GSMieee, 
        \RGnospace$^{\RGorcid}$, 
        Zuher Jahshan,
        Robert Hanhan,
        \NVnospace$^{\NVorcid}$, 
        \MLnospace$^{\MLorcid}$, \SMieee,
        \ATnospace$^{\ATorcid}$, \Mieee, 
        and \LYnospace$^{\LYorcid}$
        
\thanks{
    
    This work was supported by the Israel Science Foundation under Grant 996/18. 
    }
\thanks{Esteban Garzón is with the \UNICALfull and also with the \BIUwEnICSfull (e-mail: esteban.garzon@unical.it).}
\thanks{Roman Golman, Natan Vinshtok-Melnik and Adam Teman are with the \BIUwEnICSfull (e-mail: roman.golman@live.biu.ac.il; vinshtn@biu.ac.il; adam.teman@biu.ac.il).}
\thanks{Zuher Jahshan, Robert Hanhan, and Leonid Yavits are with the \Technionfull (e-mail: zuherjahshan@campus.technion.ac.il; roberthanhan@campus.technion.ac.il; yavits@technion.ac.il).}
\thanks{Marco Lanuzza is with the \UNICALfull (e-mail: m.lanuzza@dimes.unical.it).}
    }

\maketitle

\begin{abstract}
We propose a novel Hamming distance tolerant content-addressable memory (HD-CAM) for energy-efficient in-memory approximate matching applications.  
HD-CAM implements approximate search using matchline charge redistribution rather than its rise or fall time, frequently employed in state-of-the-art solutions. 
HD-CAM was designed in a 65\nm 1.2\V CMOS technology and evaluated through extensive Monte Carlo simulations.
Our analysis shows that HD-CAM supports robust operation under significant process variations and changes in the design parameters, enabling a wide range of \emph{mismatch threshold}  (tolerable Hamming distance) levels and pattern lengths.     
HD-CAM was functionally evaluated for virus DNA classification, which makes HD-CAM suitable for hardware acceleration of genomic surveillance of viral outbreaks such as Covid-19 pandemics.
 
\end{abstract}


\IEEEpeerreviewmaketitle

\section{Introduction}
\label{sec:Introduction}
\IEEEPARstart{C}{ontent-addressable} memories (CAMs) offer outstanding performance in applications where high-speed searching is critical~\cite{pagiamtzisCAM2006,yangCAM2011}.
In addition to well-studied applications, such as network routers, digital signal processing, analytics, and reconfigurable computing~\cite{pagiamtzisCAM2006,karamCAM2015}, CAMs can be used in variety of emerging compare-intensive big data workloads~\cite{li2019pim}, machine learning applications~\cite{mustafa2020RAMANN,Imani2016ACAM}, as well as genomic analysis ~\cite{kaplan2017resistive,taha2020approximate,kaplan2020bioseal}. 
In particular, genomic analysis, which has experienced exponential growth of data in recent years~\cite{stephens2015big}, is an active research field and the basis for different kinds of applications, such as monitoring environmental ecosystems, sustainable agriculture, Earth's environment monitoring, and personalized healthcare~\cite{glasl2019microbial,singh2020crop,zhang2020bayesian,alser2020accelerating}. Many of the those applications benefit from approximate rather than exact search, where a certain \hd, i.e., several mismatching characters between a query pattern and the dataset stored in CAM, is tolerated.



This work proposes a novel Hamming distance tolerant CAM (HD-CAM), designed to perform exact and approximate matching, capable of tolerating very large Hamming distances (e.g., 60\% of the pattern length).
Our design is based on the observation that if every mismatching bit results in a certain constant electrical charge reduction on a precharged matchline, then the total matchline voltage drop is proportional to the Hamming distance between the query pattern and a data word.
HD-CAM exploits the charge redistribution of the matchline as a measure of Hamming distance.
This is the main contribution of our proposal compared to state-of-the-art approximate CAM designs, that use the matchline rise (charge) or fall (discharge) time as an equivalence of Hamming distance~\cite{Bui2010Low,Rahimi2015Approximate,imani2017exploring}.
Another major contribution of HD-CAM is the ability to tolerate, and differentiate between patterns with, very large Hamming distances, as detailed below. 

\blue{HD-CAM applications include text processing~\cite{Ranger2007textprocessing}, DNA classification, DNA read mapping and several other genomic analysis workloads~\cite{alser2020accelerating,kaplan2018rassa}, ECC-enabled fault-tolerant CAM, as well as any other workload that requires approximate rather than exact search.}   


We performed a comprehensive design space exploration, evaluating our design in different process corners through extensive Monte-Carlo simulations.
Our study was carried out at the circuit-level using a commercial 65\nm 1.2\V CMOS technology with Cadence Spectre.
Circuit simulation results were applied to testing HD-CAM as a real-time DNA classifier, using \covid and several other virus DNA from the \ncbi online datasets~\cite{ncbi2021}.
To evaluate HD-CAM performance, we use sensitivity and specificity (defined in Section IV.A and Section V) as figures of merit. 

\begin{figure*}[t]
	\centering
	\includegraphics[width=2\columnwidth,keepaspectratio]{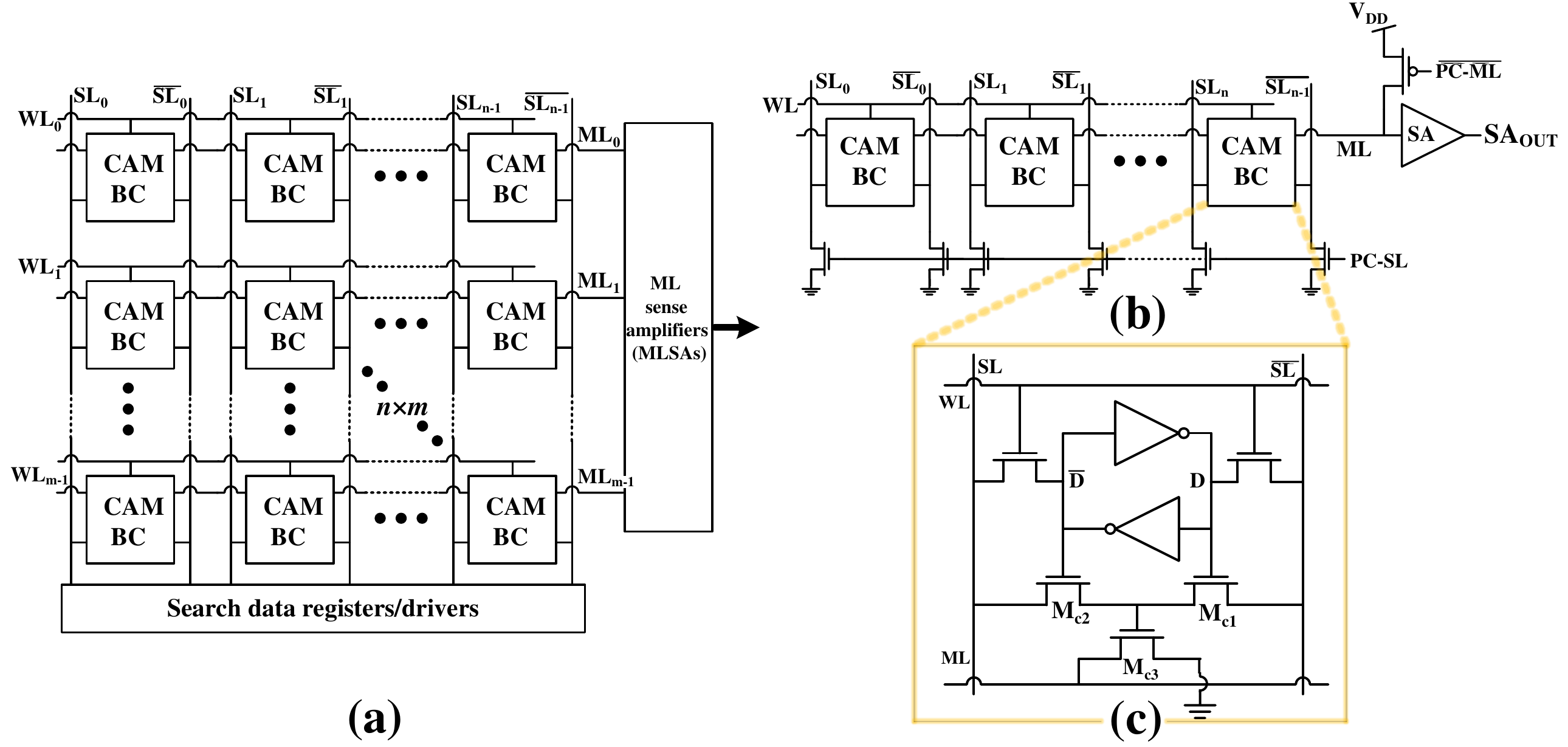}
	\caption{(a) Reference architecture for a $n$\Xnospace $m$ content-addressable memory (CAM) array. (b) CAM word of $n$-bits considering a typical pre-charge (PC) circuitry and sense amplifier (SA). (c) NOR-type CAM bitcell (BC).}
	\label{fig:camarchitecture}
\end{figure*}

\blue{To summarize, our work provides the following contributions:
\begin{itemize}
 \item HD-CAM, an approximate search CAM, that uses matchline charge redistribution as a measure of Hamming distance;
 \item HD-CAM tolerates, and differentiates between patterns with, very large Hamming distances with very high sensitivity;
 \item HD-CAM is relatively insensitive to sampling time variation;
 \item We comprehensively evaluate our design using commercial 65\nm process, covering all local variations around TT-, SS-, and FF-corners, as well as susceptibility to variations in design parameters.
\end{itemize}}

To the best of our knowledge, the HD-CAM represents the first design that can carry out approximate search, while tolerating very large Hamming distances. Moreover, it does not require data transformation such as error correction codes~\cite{Pagiamtzis2006Soft,Krishnan2009ecc} or local sensitivity hashing~\cite{ni2019ferroelectric,riazi2017camsure}.


The rest of the paper is organized as follows:
\secref{sec:Background} presents the background of our work;
\secref{sec:Proposed BC for approximate matching} discusses the proposed HD-CAM design and operation;
\secref{sec:Evaluation} details evaluation and design space exploration, while
\secref{sec:resultsandapplication} shows how HD-CAM can be used for virus DNA classification, and discusses the results;
Finally, \secref{sec:conclusions} concludes our work and presents ideas for future research.

\section{Background \& Motivation}
\label{sec:Background}
\subsection{Conventional Content-Addressable Memory (CAM)}
\label{subsec:CAM}
\figref{fig:camarchitecture}(a) shows the architecture of a conventional  $n\times m$ CAM ($n$ being the number of rows and $m$ the number of columns). 
It allows comparing a query pattern to the data stored in the bitcells. 
Each word stored in the CAM row has its own \ml, which is connected to a \sa. A pair of \sls, i.e., \sl and \slbar, are connected to all the bitcells belonging to a column.
An $n$-bit CAM word is shown in \figref{fig:camarchitecture}(b), where the \pc transistors (i.e., \pcml and \pcsl) are used to pre-discharge/charge the \sls /\ml. 
The \mlsa is used to sense the state of the \ml.

A typical NOR-type CAM bitcell is illustrated in \figref{fig:camarchitecture}(c)\footnote{It is worth mentioning that other CAM bitcell configurations also exist~\cite{pagiamtzisCAM2006}, but the remainder of this work only focuses on the NOR-type bitcell.}. 
It is based on a pair of cross-coupled inverters for storing the data. 
The bitcell is accessed for write and read similarly to a standard \sixtsram cell, by using the \wl to enable the row access, and driving \sl and \slbar to opposite values for write, or pre-charging them for read.
The associative search operation is implemented using the \McOne-\McThree transistors. 
At first, the \sls should be pre-discharged (i.e., pulled to ground), thus avoiding any possible \ml discharge. 
While keeping the \sls discharged, the \ml is pre-charged to the \vdd voltage level.
Then, the search word is loaded onto the \sls, and the \PCMLnospace transistor is turned off (i.e., \PCML = \vdd).
If the value stored in the cell matches the value on the \sls (i.e., if the \sl matches \data), \McOne and \McTwo keep the gate of \McThree low, cutting off the ML discharge path. 
In consequence, the ML remains high, which represents a match. 
On the contrary, when the \sl value differs from the value in the storage cell, \McThree turns on and discharges the \ml, which yields a mismatch.
When the entire $n$-bit word is considered (see \figref{fig:camarchitecture}(b)), the \ml will remain high only in the case that all the storage cells match the search pattern, resulting in a word match.
Conversely, a single bit mismatch is enough to  discharge the \ml, resulting in a word mismatch.

In this work, we modify the NOR-type CAM bitcell to support approximate matching, as presented hereafter.

\begin{figure}
	\centering
	\includegraphics[width=0.97\columnwidth,keepaspectratio]{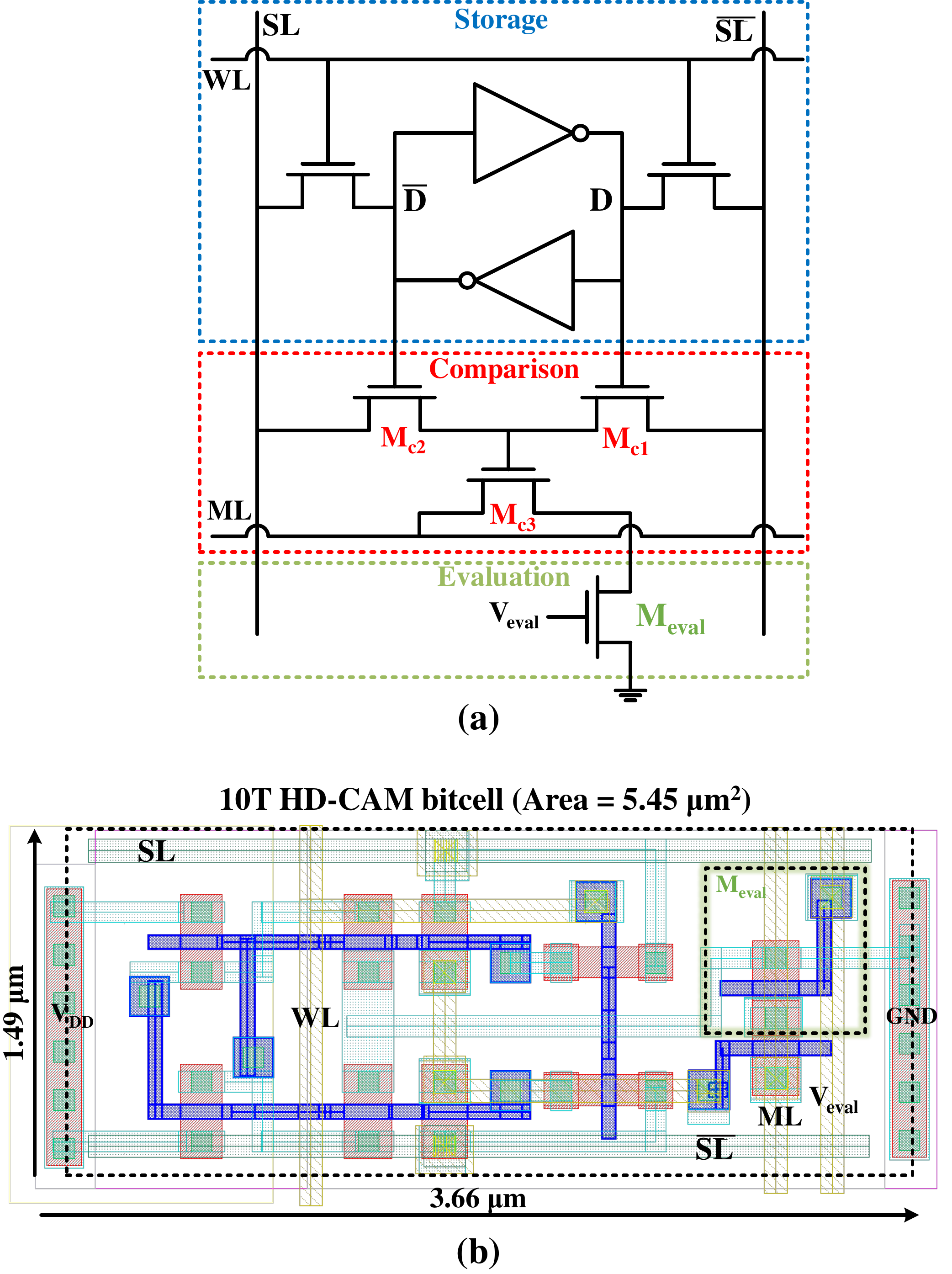}
	\caption{(a) Schematic and (b) Layout of the proposed HD-CAM bitcell.}
	\label{fig:acambitcell}
\end{figure}

\subsection{\black{Approximate Content-Addressable Memory}}
\label{subsec:HD-CAM Related Work}
Many ternary and binary NOR- and NAND-based CAM cell designs have been proposed in recent years, including CMOS-based~\cite{chang2008hybrid,kuo2003design,do2012highspeed,do2013design,sethi2017design,agarwal2011cam,jothi2018design,hussain2018match,prasanth2020high,mishra2016nineT,Arsovski2018twophase,Chan2018Reconfigurable,Yavits2013computer,Dong2018Searching,Zackriya2016Precharge}, as well as emerging memory based~\cite{Ramanathan2020Monolithic,Yavits2021GIRAF,Yavits2015Resistive,Kaplan2018PRINS,Kaplan2021Resistive} solutions.
Several CAM designs offer soft-error tolerance using error correcting coding (which requires memory redundancy) and replacing the matchline sense amplifier with an analog comparator~\cite{Pagiamtzis2006Soft,Krishnan2009ecc}.
Those designs typically tolerate only a limited Hamming distance (1-4 bits). 
Another class of approximate search CAMs uses local sensitivity hashing of stored data and query patterns~\cite{ni2019ferroelectric,riazi2017camsure}. 
While such schemes potentially tolerate large Hamming distances, they require hashing of data prior to storage and search.
Additionally, large Hamming distance does not always result in low similarity of hashed data sketches~\cite{marccais2019locality}, which leads to false negative results and hence lower sensitivity.
A CAM for minimum Hamming distance search that uses digital circuitry for bit comparison, as well as winner-take-all functionality is proposed in~\cite{Mattausch2002Compact}.
\blue{Several emerging memory (memristor crossbar) based designs for Hamming distance approximation have also been proposed~\cite{zhu2013Hamming,taha2020approximate}.}
 \blue{NCAM \cite{castaneda2019ppac} uses near-memory logic to calculate the sum of squares of data word differences (which measures the similarity between data vectors).
 PPAC \cite{del2015ncam} calculates Hamming similarity by performing a population count, by tallying the number of
ones over all XNOR outputs of the CAM bitcells of a word.}


A variety of approximate search CAM designs use timing (i.e., score signal delay, or the speed of the matchline discharge) as a measure of Hamming distance.
A Hamming distance search CAM, where the score signal is delayed every time a bit mismatch occurs, is proposed in~\cite{Bui2010Low}.
In this design, the delay of the score signal is proportional to the Hamming distance between the search and stored patterns.
In the approximate search enabled CAM for energy efficient GPUs, proposed in~\cite{Rahimi2015Approximate}, a small Hamming distance ($\leq2$ bits) is tolerated through meticulous timing of the matchline discharge.
In~\cite{imani2017exploring}, Hamming distance of ($\leq4$ bits) is tolerated by using delay lines at the clock inputs of four separate sense amplifiers on each matchline. 
These tunable sampling time techniques require very precise device and circuit sizing, and suffer from false negatives (false mismatches) as well as false positives (multiple false matches)~\cite{Rahimi2015Approximate}, leading to limited efficiency of the approximate search technique. Tuning the sampling time is a complex task, which would require almost perfect skew balancing between all \ml timing circuits and would be very sensitive to jitter. These issues are exacerbated by process variations.


\subsection{DNA Classification Using Approximate Search CAM}
\label{subsec:DNA class. approx.CAM}
DNA sequencing
is used for genomic surveillance and variant classification during the ongoing Covid-19 pandemic. 
DNA sequencing is a process of determining the bases of a DNA chain, which are referred to as Adenine (A), Guanine (G), Cytosine (C), and Thymine (T). 
Contemporary high-throughput DNA sequencers can sequence multiple DNA samples in parallel~\cite{illumina2021}.
DNA sequencing process, along with the genomic analysis, is carried out in several steps~\cite{kim2018grim}: (1) sample preparation; (2) DNA sequencing that generates multiple DNA fragments called DNA reads; and (3) DNA classification, DNA read alignment, genome assembly, variant analysis, etc.

Typically, tools like Kraken and Kraken2~\cite{wood2014kraken,wood2019improved} are used to detect and classify unknown DNA.
However, Kraken operation is based on exact matching of sequenced DNA patterns in DNA database.
Therefore, it requires relatively high coverage (high percentage of the target DNA in a sample) to perform with sufficient sensitivity. 

In this work, we propose a fast and highly sensitive approximate matching-based DNA classification scheme, implemented by HD-CAM.
Our design allows tolerating very large Hamming distances (for example, up to 60\% of the pattern length), while providing very high sensitivity and specificity, as detailed in \secref{sec:resultsandapplication}.


\begin{figure}
	\centering
	\includegraphics[width=1.0\columnwidth,keepaspectratio]{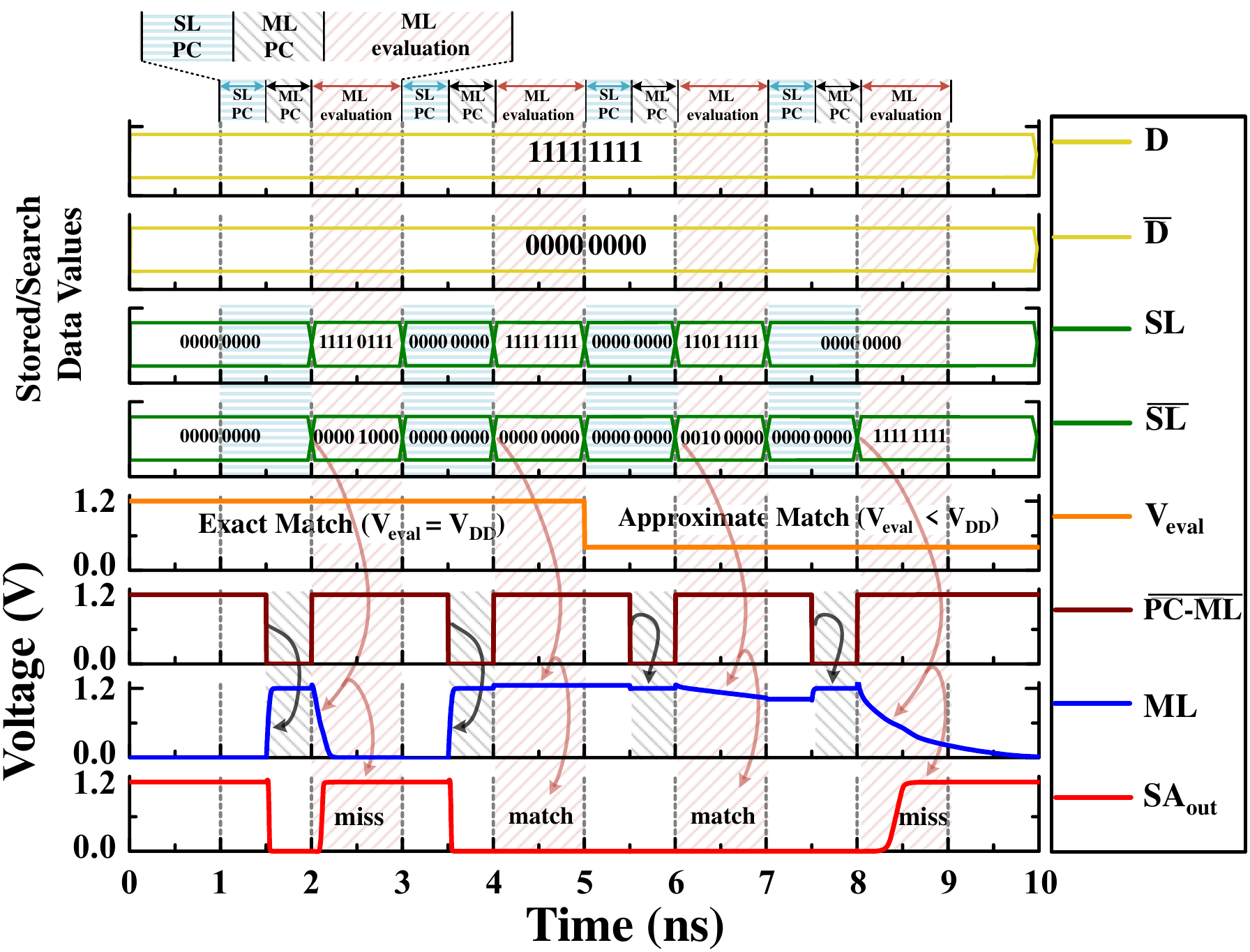}
	\caption{Timing diagram for exact and approximate matching of the proposed HD-CAM organized as a single 8-bit word. Note that the stored data, \data, is set to logical \one, and the nominal voltage is 1.2\V.}
	\label{fig:acamoperation}
\end{figure}

\section{\blue{Hamming Distance tolerant CAM (HD-CAM) design}}
\label{sec:Proposed BC for approximate matching}
\vspace{0mm}
\subsection{\blue{Bitcell design}}
\label{subsec:hdcam bitcell}
The goal of our design is providing highly confident match and mismatch for dynamically configurable (by user) mismatch threshold (i.e., the Hamming distance, or the number of mismatching bits that can be tolerated), while making the proposed HD-CAM resilient to the process and design parameters variation. 
In other words, we want to ensure a definite match if the number of mismatching bits in an HD-CAM row is $\leq$ \emph{k}, and a certain mismatch if the number of mismatching bits in an HD-CAM row is $\geq$ \emph{l}, where \emph{k} and \emph{l} are integer numbers 
and \emph{k}$<$\emph{l}.
Throughout the paper, we refer to the region between \emph{k} and \emph{l} as \emph{uncertainty region}, where a false match or a false mismatch may occur.  

\figref{fig:acambitcell}(a) shows the schematic of the HD-CAM bitcell, which is capable of performing exact as well as approximate search operations. 
The HD-CAM cell is based on the NOR-type CAM bitcell of \figref{fig:camarchitecture}(c) with the addition of an evaluation transistor (\Meval) that is used to control the \ml discharge rate, based on the level of the evaluation voltage (\Veval).
When the \Meval transistor is driven by full voltage level (i.e., \Veval = \vdd), a conventional exact match CAM operation is performed. 
Approximate matching is enabled when \Veval $<$ \vdd. 
The layout of the HD-CAM bitcell is shown in \figref{fig:acambitcell}(b).
Note, the HD-CAM cell is laid out using Metal-1 to Metal-3 only, and can be further improved by allowing higher Metal layers.

\figref{fig:acamoperation} shows the timing diagram of the HD-CAM, organized as a single 8-bit word, for the cases where \Veval= \vdd (exact match) and \Veval $<$ \vdd (approximate match).
In the simulation, 
each cycle is 2\ns long made up of 1\ns of pre-charge and 1\ns of evaluation time (\teval).
For both cases, the operating principle is the same: at first, \sl and \slbar are driven low. 
Then, the \ml is pre-charged to \vdd and subsequently, the search pattern is loaded to the \sls for evaluating the \ml. 
If the \ml stays above  a certain evaluation threshold level (defined in \secref{subsec:SA}), we achieve a match; otherwise there is a mismatch.
For circuit simulation purposes, we pre-loaded the 8-bit HD-CAM word with \data = \one (see the \data signal in \figref{fig:acamoperation}); then, in each subsequent cycle we applied a new query pattern to the \sls.
In the case of the exact search operation (see \figref{fig:acamoperation}, 0-5\ns time-frame), a single bit mismatch between \data and \sl results in a full discharge of the \ml, while a match will keep \ml high.
In the case of approximate search (see \figref{fig:acamoperation}, 5-10\ns time-frame), the \ml remains high even with a single mismatching bit. 
Finally, when several (or all, as in the example shown in \figref{fig:acamoperation}, 8-10\ns time-frame) bits are different, the ML fully discharges.

\begin{figure}[t]
	\centering
	\includegraphics[width=1.01\columnwidth,keepaspectratio]{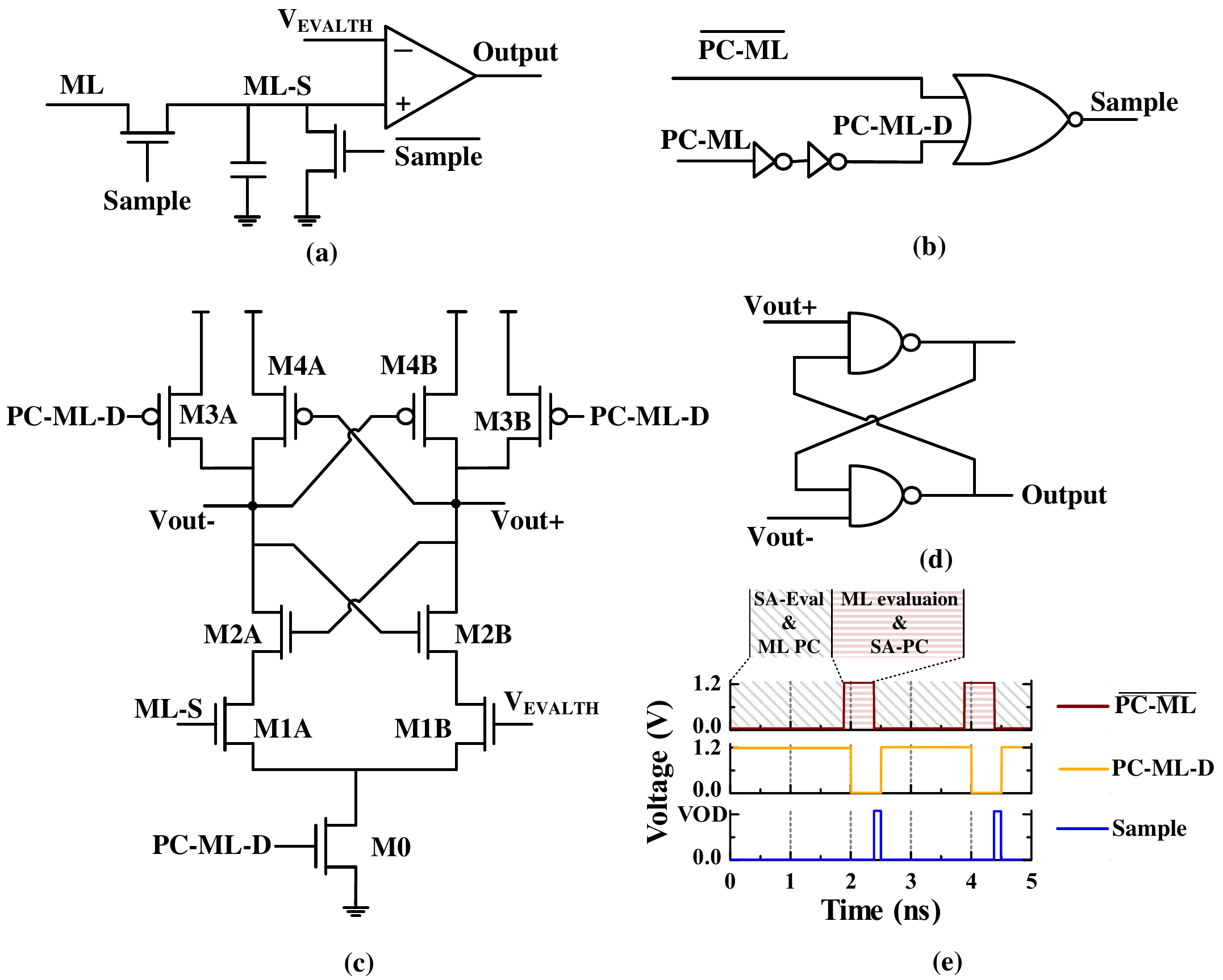}
	\caption{MLSA Structure, Timing Circuit and Waveforms.}
	\label{fig:MLSAStruct}
\end{figure}

Unlike the exact matching operation, the approximate matching task depends on the conductivity of the \Meval transistor, which plays a major role.
When \Veval is low enough, a single mismatching cell is not able to discharge the ML capacitance during the evaluation time (see \figref{fig:acamoperation}, 6-7\ns time-frame).
\blue{This leads to longer evaluation time in contrast to the exact match operation (see 2-3\ns and 8-9\ns time-frame in \figref{fig:acamoperation}) due to the inherently lower conductivity of the \Meval transistor when the asserted \Veval is lower than \vdd, causing the \ml to discharge slowly.}
When taking into account HD-CAM arrays with longer words (patterns), e.g., 128-bit or 256-bit, the inherently larger \ml capacitance will impact the approximate match operation, and several mismatching bitcells may be required to discharge the \ml.

\blue{HD-CAM exhibits certain delay overhead compared to conventional NOR CAM schemes with no evaluation transistor.
As we show in \secref{sec:Evaluation}, such delay overhead is the direct result of HD-CAM ability to enable very large mismatch threshold, and to maintain very high efficiency across the entire range of mismatch threshold levels. 
}

\begin{table*} 
    \centering
    \captionsetup{justification=centering}
    \caption{CAM vs HD-CAM comparison \black{at \tt corner}}
        \includegraphics[width=0.84\textwidth]{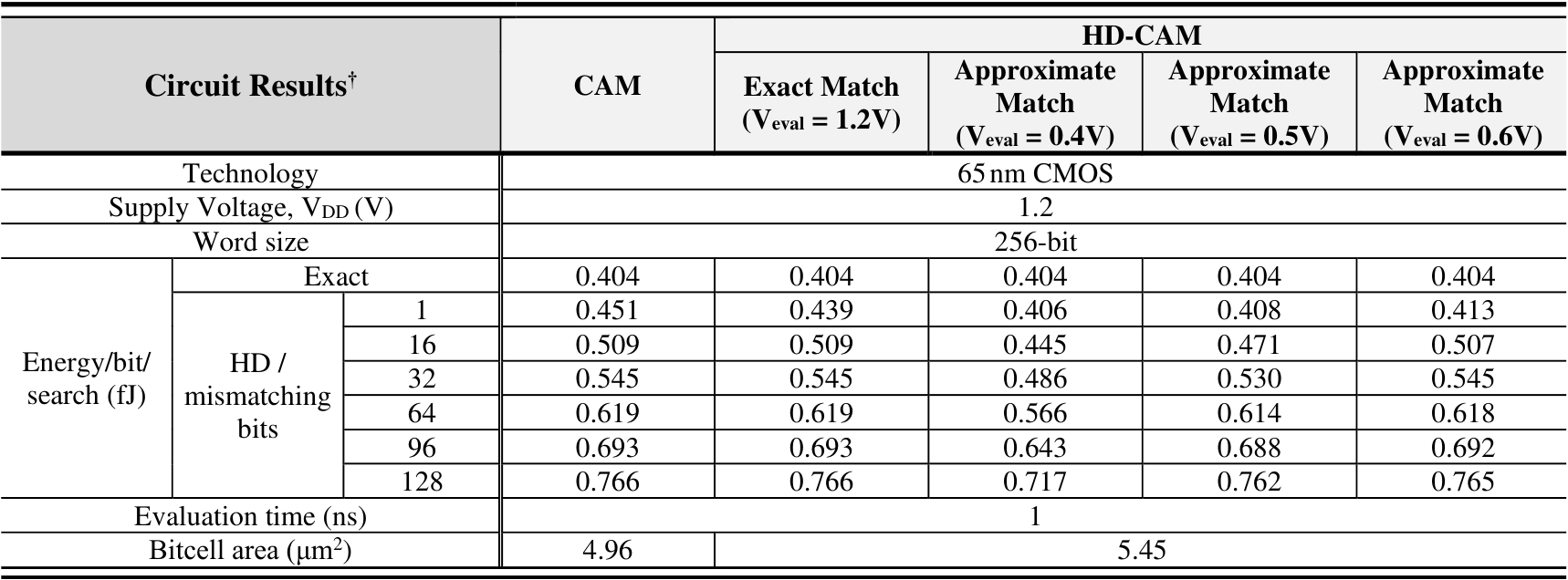} 
    \label{tab:Table 2}
\end{table*}
\subsection{\blue{Sense Amplifier Design}}
\label{subsec:SA}

\blue{The matchline sense amplifier (\mlsa) is based on a classical StrongARM design \cite{SA}.
The \ml sensing design and circuitry is shown in \figref{fig:MLSAStruct}, as follows: a Sample and Hold circuit (S\&H) (\figref{fig:MLSAStruct}(a)), a timing unit (\figref{fig:MLSAStruct}(b)), a cross-coupled amplifier (\figref{fig:MLSAStruct}(c)) which, in an open loop configuration works as a comparator, and a RS latch (\figref{fig:MLSAStruct}(d)).
While ML is being evaluated, the cross-coupled amplifier is in a PC phase dictated by M3A and M3B transistors controlled by a delayed PC-ML (i.e., PC-ML-D), where the outputs are charged to \vdd.
Upon completion of the evaluation phase, a pulse (Sample) triggers the charging of a small capacitor to the ML voltage - labeled ML-Sampled (ML-S). The comparator evaluates the ML-S compared to an evaluation threshold (\Vevalth), which is the tolerance reference voltage, yielding a match or mismatch when the \ml voltage is above or below \Vevalth, respectively.
The result is fed into the RS latch, which compensates for metastable behavior in the \mlsa PC phase. Sampling the value of ML onto a capacitor frees the mutual dependence of the evaluation time of the MLSA response and the PC of the ML for the next evaluation. This allows the creation of two interlocked cycles - while ML is precharged, the S\&H result is being evaluated, and while the SA is being precharged, the ML is being evaluated. This results in zero-downtime of the system, and every PC cycle has a corresponding evaluation cycle done in parallel as shown in Fig. \ref{fig:MLSAStruct}(e). In addition to the sampling device, an inverse device exists, in order to discharge the ML-S node for the next sample.}




\section{Evaluation and Design Space Exploration}
\label{sec:Evaluation}
\blue{HD-CAM evaluation and design space exploration are carried out using extensive \mc simulations.
We employ Cadence Spectre, using transistor models provided by a commercial 65\nm CMOS technology featuring a nominal supply voltage \vdd of 1.2\V. }

\blue{HD-CAM design space is defined by the tuple [\teval, \Veval, $V_{\text{evalth}}$], which we interchangeably refer to as design space variables or parameters. In the following subsections, we show how adjusting those parameters allow us to tune the desired mismatch threshold as well as HD-CAM efficiency levels for different process corners.}

\subsection{\blue{Energy consumption and silicon area}}
\label{subsec:energy and area}
\tblref{tab:Table 2} provides a comparison, at circuit-level, between a conventional CAM and the proposed HD-CAM in terms of energy \blue{consumption-per-bit and bitcell area.}
We assume \black{a memory word of 256-bits with} a single operation cycle of 2\ns (\pc + evaluation time, as shown previously in \figref{fig:acamoperation}).
Additionally, for HD-CAM we have examined three \Veval values: 0.4\V, 0.5\V, and 0.6\V.
For exact match operations, both the HD-CAM (with \Veval\unskip=1.2\V) and conventional CAM have the same energy consumption.
However, in comparison to conventional CAM, HD-CAM consumes less energy per bit when performing an approximate search operation, due to the partial (rather than complete) discharge of the matchline.
The results show that for the various \hd levels, energy consumption per bit grows as \Veval increases. 
\black{
Bitcell area comparison is shown in \tblref{tab:Table 2}. HD-CAM employs an additional transistor (see \figref{fig:acambitcell}(b)) leading to a bitcell area overhead of about 10\%.
}

\begin{figure*}[h!]
	\centering
	\includegraphics[width=2.05\columnwidth,keepaspectratio]{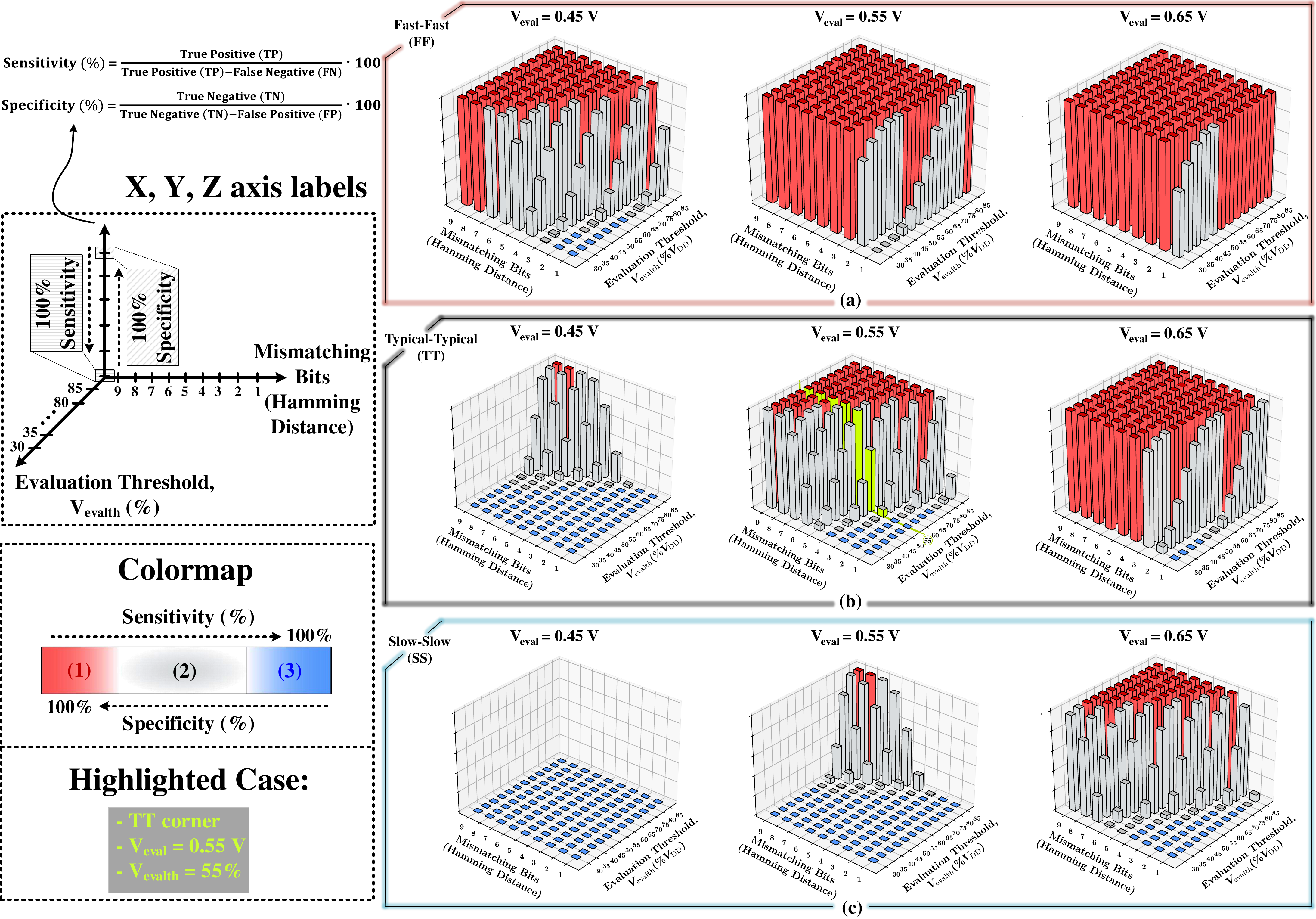}
	\caption{HD-CAM susceptibility to process variation: HD-CAM sensitivity/specificity vs. \Vevalth (represented in \% of \vdd) and vs. Hamming distance (between the query pattern and a stored data word), for three \Veval levels and for (1) Fast-Fast (FF), (b) Typical-Typical (TT), and (c) Slow-Slow (SS) corners, based on 1,000 Monte Carlo simulations. The evaluation time, \teval is 1\ns.}
	\label{fig:approximatesensitivity}
\end{figure*}


\subsection{\blue{Figures of merit for HD-CAM approximate search efficiency}}
\label{subsec:figures of merit}
\blue{The first figure of merit is \emph{Sensitivity}, which measures the probability of correctly detecting the similarity existing between the query pattern and any of the data words stored in HD-CAM. It is defined as:
\begin{equation}
Sensitivity = \frac{TP}{(TP+FN)},
\end{equation}
where $TP$ and $FN$ are true positive and false negative results, respectively.}
\blue{True positive (TP) result is obtained when a compare of two patterns with Hamming distance \textbf{below} the mismatch threshold results in a \textbf{match};}
\blue{False negative (FN) result is obtained when a compare of two patterns with Hamming distance \textbf{below} the mismatch threshold results in a \textbf{mismatch}.}
\blue{More plainly, sensitivity measures the probability of correctly detecting the similarity that actually exists between the query pattern and any of the data words stored in HD-CAM}. 

\blue{The second figure of merit is \emph{Specificity}, which measures the probability of correctly rejecting a query pattern which is not sufficiently similar to any of the data words stored in HD-CAM. It is defined as:
\begin{equation}
Specificity = \frac{TN}{(TN+FP)},
\end{equation}
where $TN$ and $FP$ are true negative and false positive results, respectively.}
\blue{True negative (TN) result is obtained when a compare of two patterns with Hamming distance \textbf{above} the mismatch threshold results in a \textbf{mismatch};}
\blue{False positive (FN) result is obtained when a compare of two patterns with Hamming distance \textbf{above} the mismatch threshold results in a \textbf{match}.}
\blue{Specificity measures the probability of correctly rejecting a query pattern which is not sufficiently similar to any of the data words stored in HD-CAM.}

\subsection{\blue{Susceptibility to process variation}}
\label{subsec:suscept PV}
We apply extensive \mc simulations to examine local variations (i.e., mismatches\footnote{This is a process variation term, not to be confused with HD-CAM mismatch}) around \ff, \tt, and \ss corners.
In order to assess the entire design space of HD-CAM, three main design parameters that affect the \ml signal are examined:
the \Veval voltage applied at the gate of the \Meval transistor (see \figref{fig:acambitcell}), the evaluation time \teval, and the evaluation threshold \Vevalth.
\blue{Each combination of [\teval, \Veval, $V_{\text{evalth}}$] corresponds to an individual mismatch threshold level. In other words, mismatch threshold can be adjusted by changing the values of [\teval, \Veval, $V_{\text{evalth}}$]. 
}

We simulate the \blue{HD-CAM approximate search sensitivity and specificity} 
for different process corners and different values of \Vevalth and \Veval, as presented in \figref{fig:approximatesensitivity}.
Here, \teval is set at 1\ns (following the pre-charge stage).
Lastly, we vary the Hamming distance, between the query pattern and the stored data word, from 1 bit to 9 bits.


\blue{Due to process variations, different HD-CAM memory rows may require different numbers of mismatching bits to keep the \ml above the \Vevalth level (producing a match), as well as to discharge the \ml below the \Vevalth level (yielding a mismatch).
In other words, HD-CAM sensitivity and specificity are affected by process variations.
We differentiate between three sensitivity/specificity regions (see color-map of \figref{fig:approximatesensitivity}): (1) 100\% specificity, i.e., mismatches are always true; (2) an uncertainty region with sensitivity and specificity both below 100\% (meaning matches and mismatches could be true and false); and (3) 100\% sensitivity, i.e., matches are always true.
Overall, \figref{fig:approximatesensitivity}(a-c) show that uncertainty regions are relatively small, whereas 100\% sensitivity and 100\% specificity regions (plateaus) are accordingly very large.}


\blue{Some applications, such as DNA classification presented in Section V, tolerate uncertainty (i.e., sensitivity and specificity under 100\%) due to their intrinsic inexact nature.
Other applications, that require 100\% sensitivity and 100\% specificity, may rely on advanced coding techniques that ensure that any two valid codewords should have Hamming distance larger than the worst case uncertainty region (2). In such case, a valid codeword is always differentiated from any other valid codeword with probability of 100\% (due to 100\% specificity), while small changes (caused for example by soft errors) that fall into the blue region (3) are tolerated with 100\% sensitivity.}     

\blue{Consider the following example. Suppose a certain application requires the mismatch threshold of 2 bits with 100\% sensitivity. For TT corner, this requirement is satisfied by setting \Veval at 0.55V and \Vevalth at 55\% \vdd. This scenario is marked in fluorescent yellow in \figref{fig:approximatesensitivity}(b). However, if the HD-CAM chip targeted for this application is manufactured in FF corner, the required mismatch threshold level can still be maintained by reducing \Veval to 0.45V and keeping the \Vevalth under 45\% \vdd. Similarly, if the HD-CAM chip is manufactured in SS corner, the required mismatch threshold can be sustained by increasing \Veval to 0.65V and keeping the \Vevalth under 70\% \vdd. This example shows that HD-CAM can successfully endure significant process variation and keep the target mismatch threshold by dynamically adjusting the [\teval, \Veval, $V_{\text{evalth}}$] tuple.}


\begin{figure}
	\centering
	\includegraphics[width=1\columnwidth,keepaspectratio]{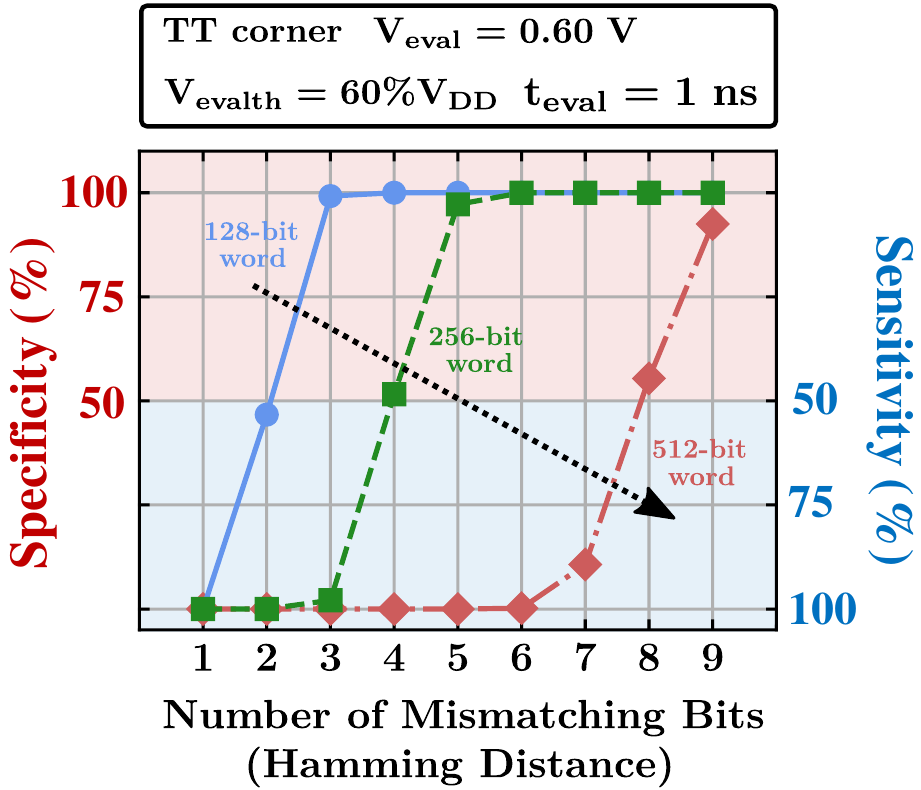}
	\caption{\blue{HD-CAM sensitivity and specificity} levels for different word sizes (pattern lengths) for \tt-corner, \Veval 0.6\V, \Vevalth 0.60\%\vdd, and \teval 1\ns.
	}
	\label{fig:matchconfidence}
\end{figure}

\subsection{\blue{Susceptibility to pattern length}}
\label{subsec:suscep pattern len}
\blue{State  of  the  art  approximate  search  techniques  that  use tunable  sampling  time  to  measure  similarity, support limited pattern length, typically up to 64 or 96 bits~\cite{Rahimi2015Approximate}. While sufficient in several applications, this might be quite prohibitive in genomic analysis applications.}

We extend our analysis on the HD-CAM to evaluate its behavior for different pattern length (memory word sizes): 128-bit, 256-bit, and 512-bit.
\blue{We assume the mismatch threshold of zero and conduct} \black{1000} \mc simulations for Hamming distance ranging from 1 to 9 bits.
\figref{fig:matchconfidence} presents the \blue{sensitivity and specificity} vs. \hd, for different  word lengths, for the \tt-corner with \Veval\unskip=0.60\V, \Vevalth\unskip=0.6$\cdot$\vdd, and \teval\unskip=1\ns.

\blue{
We observe that for the fixed values of \teval, \Veval and \Vevalth, larger memory words allow higher mismatch threshold levels at the cost of slightly wider uncertainty regions (regions where the sensitivity and specificity are lower than 100\%)}.

\begin{figure*}[h!]
	\centering
	\includegraphics[width=2.05\columnwidth,keepaspectratio]{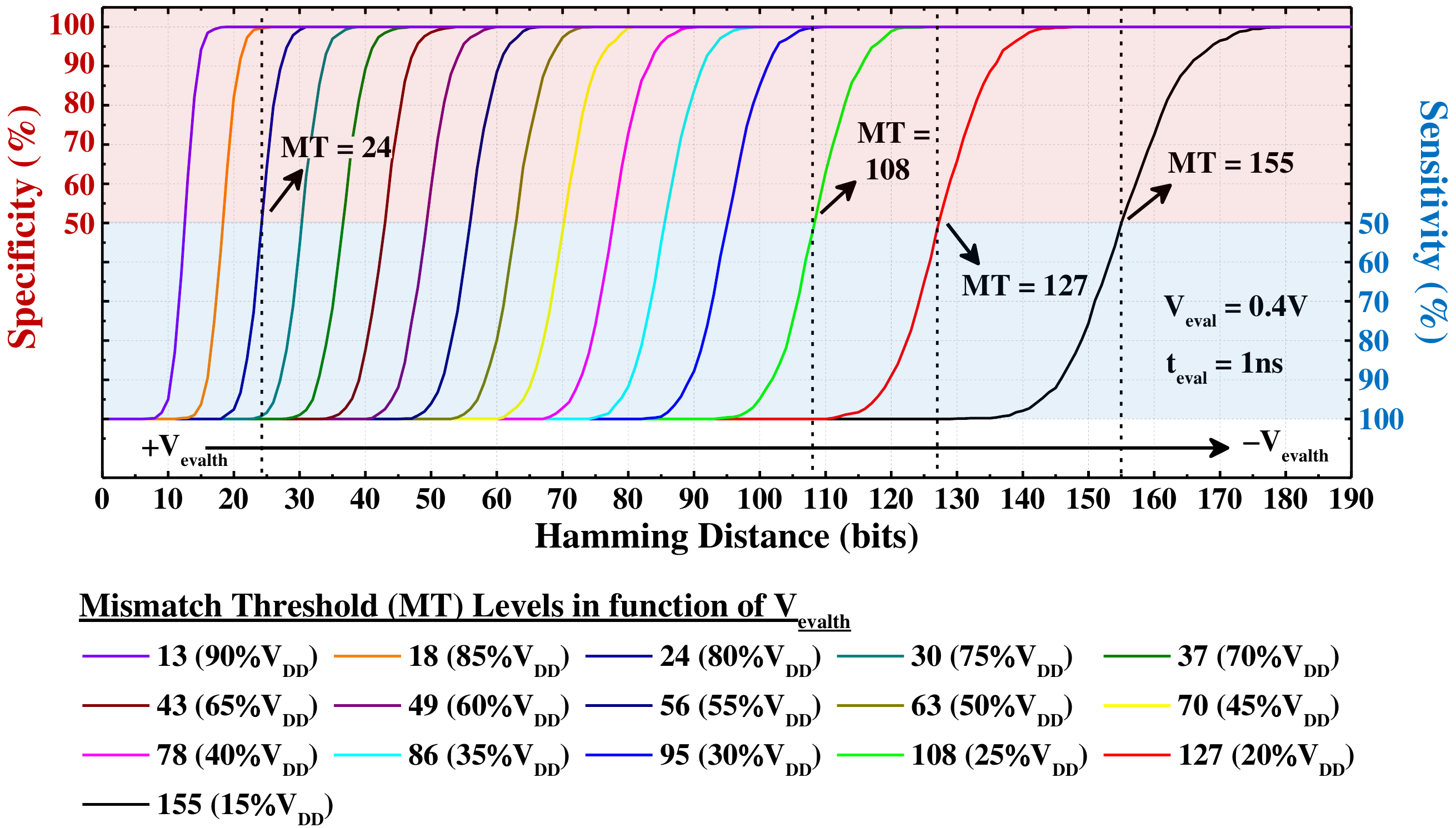}
	\caption{Sensitivity and Specificity vs. Hamming distance for different mismatch threshold (MT) levels. Sensitivity (specificity) is the lowest when Hamming distance equals the mismatch threshold but quickly grows as the Hamming distance drops below (rises above) mismatch threshold}
	\label{fig:sensitivity_specificity}
\end{figure*}


\begin{figure}[ht!]
	\centering
	\includegraphics[width=1\columnwidth,keepaspectratio]{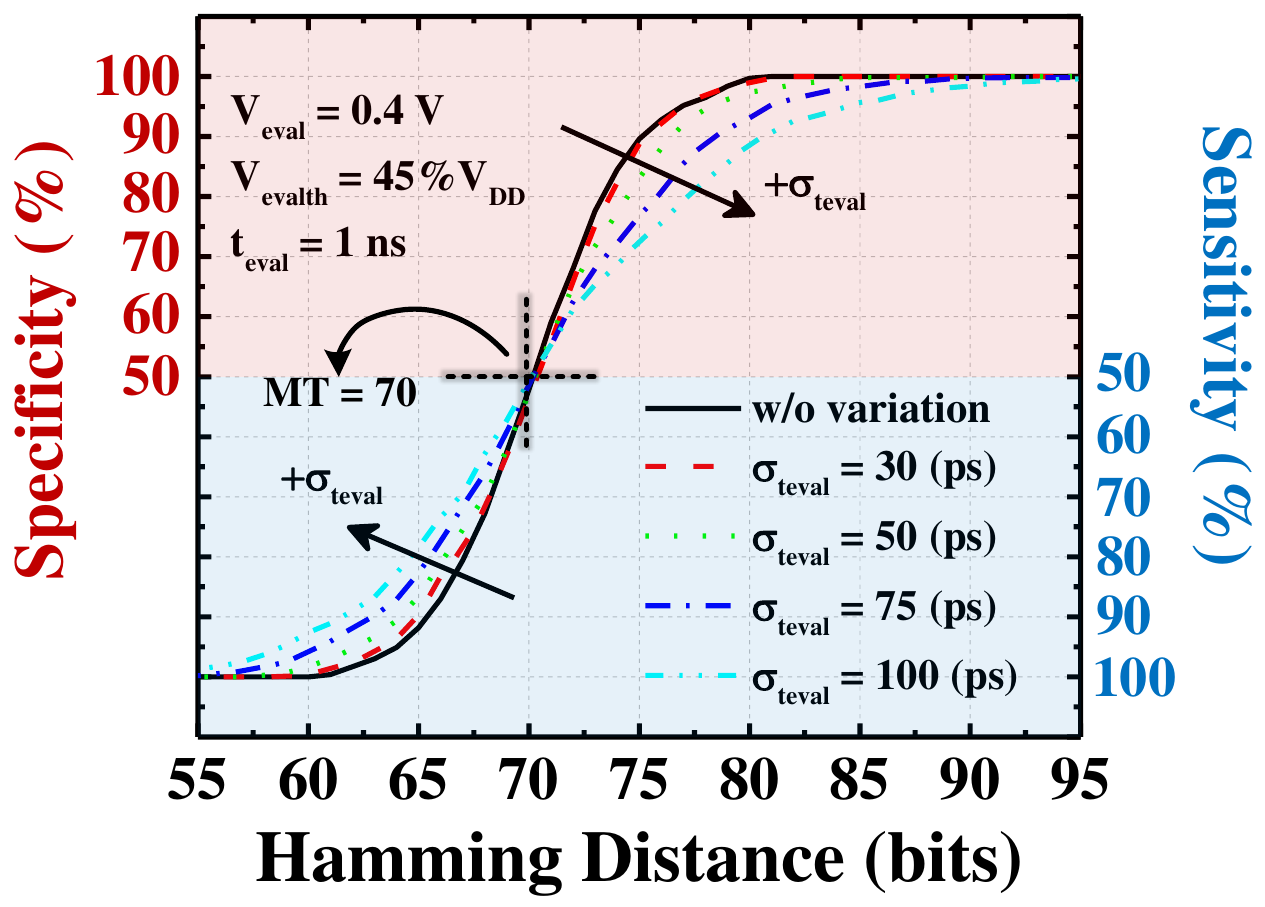}
	\caption{HD-CAM sensitivity to the sampling time \teval variation. \teval is normally distributed with mean of 1\ns and standard deviations of 30\ps, 50\ps, 75\ps, and 100\ps.}
	\label{fig:teval drift}
\end{figure}

\subsection{\blue{Susceptibility to mismatch threshold level}}
\label{subsec:suscept MT}
\blue{Mismatch threshold defines the tolerable Hamming distance. State of the art approximate search techniques that use tunable sampling time to measure similarity, are limited to tolerating very small Hamming distances (typically, 1-4 bits), and demonstrate limited sensitivity and precision~\cite{Rahimi2015Approximate}. While suitable for some applications, it might be quite prohibitive in applications such as text processing and genomic analysis, that require similarity search with insertions and deletions. Insertions are edits in which extra character(s) or DNA basepair(s) are inserted into existing text string or DNA sequence. Deletions are edits in which character(s) or DNA basepair(s) are deleted from existing text string or DNA sequence. A single insertion or deletion may result in a very large Hamming distance, so that a query pattern with a single insertion or deletion will not match, thus reducing the sensitivity of the approximate search technique.}

\blue{The proposed HD-CAM is capable of tolerating, and differentiating between patterns with, very large Hamming distances.
\figref{fig:sensitivity_specificity} shows the sensitivity and specificity of HD-CAM for several mismatch threshold levels, spanning almost the entire pattern length.}

\blue{Based on sensitivity and specificity results of \figref{fig:sensitivity_specificity}, we can make the following observations:
\begin{itemize}
\item HD-CAM supports very large mismatch thresholds, up to 155 bits out of 256 bits, or above 60\% of the pattern length in this study (well above the typical limit of 4 bits in state of the art tunable sampling time schemes).
\item HD-CAM exhibits sensitivity (specificity), typically around 55\%-60\%, when the Hamming distance between the query pattern and a stored word is as high as \emph{mismatch threshold - 1 (+1)}  bits. 
\item  HD-CAM sensitivity (specificity) reaches 100\% relatively quickly as the distance between the query pattern and a stored word drops (rises).
\item Sensitivity and specificity rise slower as the mismatch threshold increases. However, even for mismatch threshold of 155 bits (60\% of the pattern length), the sensitivity and specificity reach 100\% when the Hamming distance between the query pattern and a stored data word rises (drops) 24 bits above (below) the mismatch threshold of 155 bits. 
\end{itemize}}

\blue{In summary, HD-CAM is not limited to tolerating of, and differentiating between patterns with, small Hamming distances, which is prohibitive in several similarity search applications. HD-CAM can extend the mismatch threshold to at least 60\% of the pattern length. Note, in this study we only change the \Vevalth to tune the mismatch threshold. Higher mismatch thresholds and narrower uncertainty regions can potentially be obtained if we additionally adjust other design parameters, such as \Veval.}  

\subsection{\blue{Susceptibility to \teval variation}}
\label{subsec:teval drift}
\blue{State  of  the  art  approximate  search  techniques  that  use tunable  sampling  time  to  measure  similarity, are strongly susceptible to time variation and jitter. In contrast, HD-CAM is specifically designed to withstand time variation.} 

\blue{In order to quantitatively evaluate the sensitivity of HD-CAM to the sampling time \teval variation, we perform 1000 \mc  simulations with random variable \teval, normally distributed with the mean of 1\ns and standard deviations of 30\ps, 50\ps, 75\ps and 100\ps.
Sensitivity and specificity of the HD-CAM approximate search under \teval variation are presented in \figref{fig:teval drift}, for a large mismatch threshold of 70 bits.}
\blue{The \mc results demonstrate that \teval variation impact is limited to a certain expansion of the uncertainty region.}

\subsection{\blue{Summary of the design space exploration}}
\label{subsec:summary design space}
\blue{We showed that by adjusting design space parameters  \teval, \Veval and \Vevalth, we are able to:} 
\begin{enumerate}
\item \blue{Successfully and efficiently calibrate the mismatch threshold, above 60\% of the pattern length and potentially higher;}

\item \blue{Ensure the sensitivity and specificity of 100\%;}

\item \blue{Successfully tolerate different process corners. For  example,  if a target  application  requires  a certain  mismatch  threshold assuming TT-corner,  but  the  specific HD-CAM chip is manufactured at FF-corner or  SS-corner, we would  be  able  to  maintain  the  required  mismatch  threshold  by adjusting \Veval and \Vevalth.}
\end{enumerate}


\section{Application \& Results}
\label{sec:resultsandapplication}
Genomic analysis can be used for  diagnosis of viral diseases alongside traditional testing methods such as \pcr ~\cite{bloom2020swab,garibyan2013research,artika2020pathogenic}.
It provides much higher sensitivity (i.e., much lower false negative rate) and precision (i.e., much lower false positive rate) than \pcr for virus detection in living organisms~\cite{bloom2020swab,Shamsi2021covid}.

Covid-19 has emerged as a worldwide pandemic, causing the loss of millions of lives and tens of trillions of dollars. 
For a rapid response to pandemic outbreak, a fast, cost-effective, and reliable pathogen classification is strongly needed~\cite{bloom2020swab,artika2020pathogenic}.
Particularly, accurate diagnostic is extremely important for the governments to establish appropriate control measurements and guidelines~\cite{armstrong2020covid}.
If genomic surveillance of the entire population combined with efficient and affordable analysis of sequenced DNA were available, this goal would be more achievable.

\subsection{\blue{HD-CAM as DNA classifier}}
\label{subsec:DNA classifier}

\begin{figure}
	\centering
	\includegraphics[width=1.00\columnwidth,keepaspectratio]{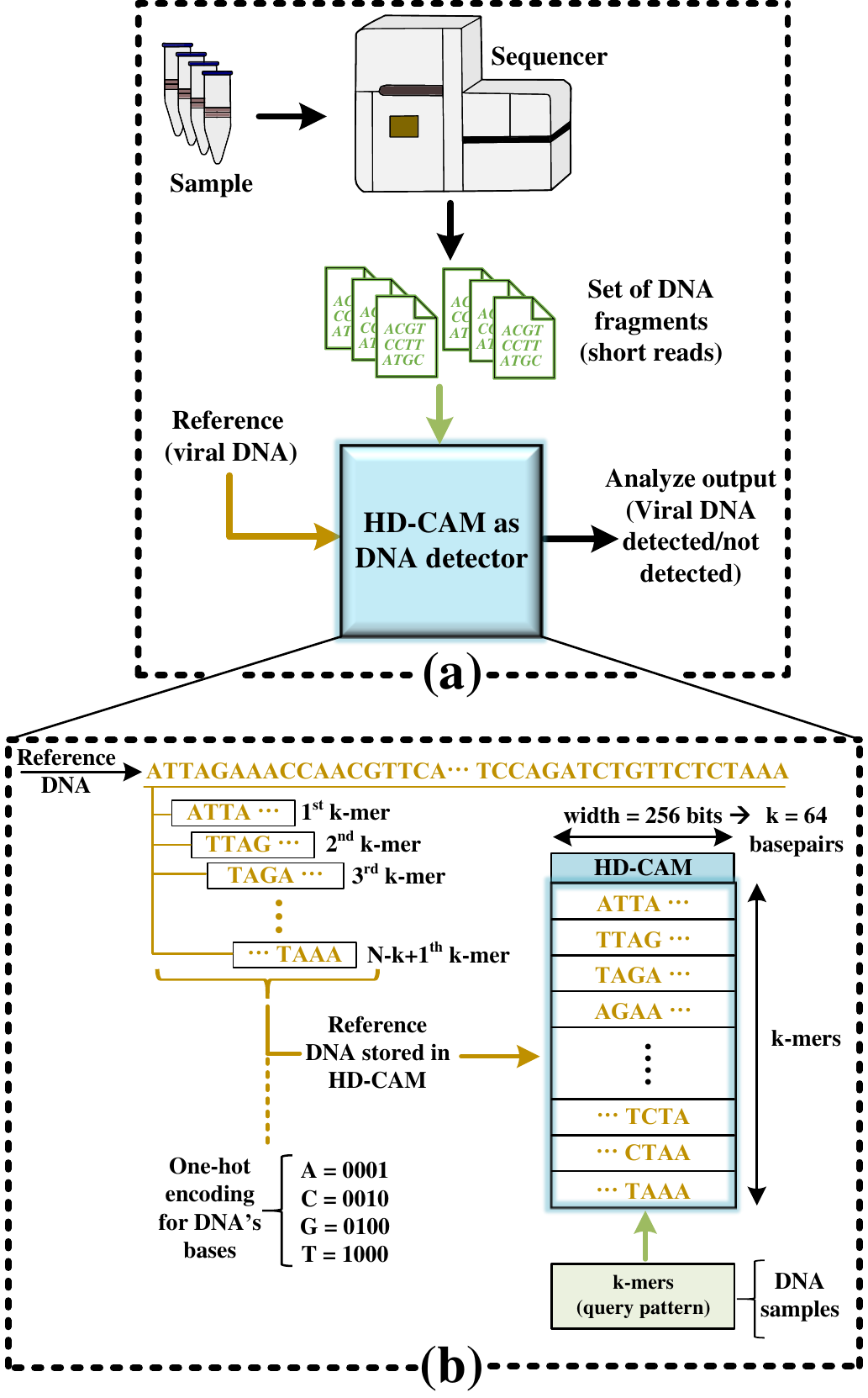}
	\caption{(a) Proposed HD-CAM as part of the general pipeline for DNA classification. (b) Construction of the reference DNA database in HD-CAM (left) and DNA classification operation (right).}
	\label{fig: HD-CAM as DNA classifier}
\end{figure}

Virus DNA classification, whose aim is to overcome the shortcomings of existing testing techniques~\cite{garibyan2013research,artika2020pathogenic}, can strongly benefit from the approximate matching capabilities of the proposed HD-CAM. 
\figref{fig: HD-CAM as DNA classifier}(a) illustrates the HD-CAM as a component in a virus DNA classification-by-sequencing pipeline, where for the sake of simplicity, intermediate steps for the DNA sample preparation and sequencing were omitted.

\blue{The DNA of the target virus (designated as the reference DNA) is known in advance, and represented by a set of short DNA fragments ($k$ basepairs long) named \kmers~\cite{ng2017dna2vec}.} 
All possible \kmers in the reference DNA are extracted as follows (refer to \figref{fig: HD-CAM as DNA classifier}(b)): the first \kmer is the reference DNA fragment from position 0 to position $k-1$, the second \kmer is the reference DNA fragment from position 1 to position $k$, and so on.
The reference DNA database is generated by storing \kmers in the HD-CAM prior to the classification operation, where a unique \kmer is stored in a separate HD-CAM row, as shown in \figref{fig: HD-CAM as DNA classifier}(b).
The number of \kmers in the HD-CAM is bounded by $N-k+1$, where $N$ is the length of the virus DNA.
In the case of \covid virus, $N$ is equal to 29,903, where some \kmers may appear more than once, therefore the actual number of \kmers in the HD-CAM could be lower.


\begin{figure*}[h!]
	\centering
	\includegraphics[width=2.0\columnwidth,keepaspectratio]{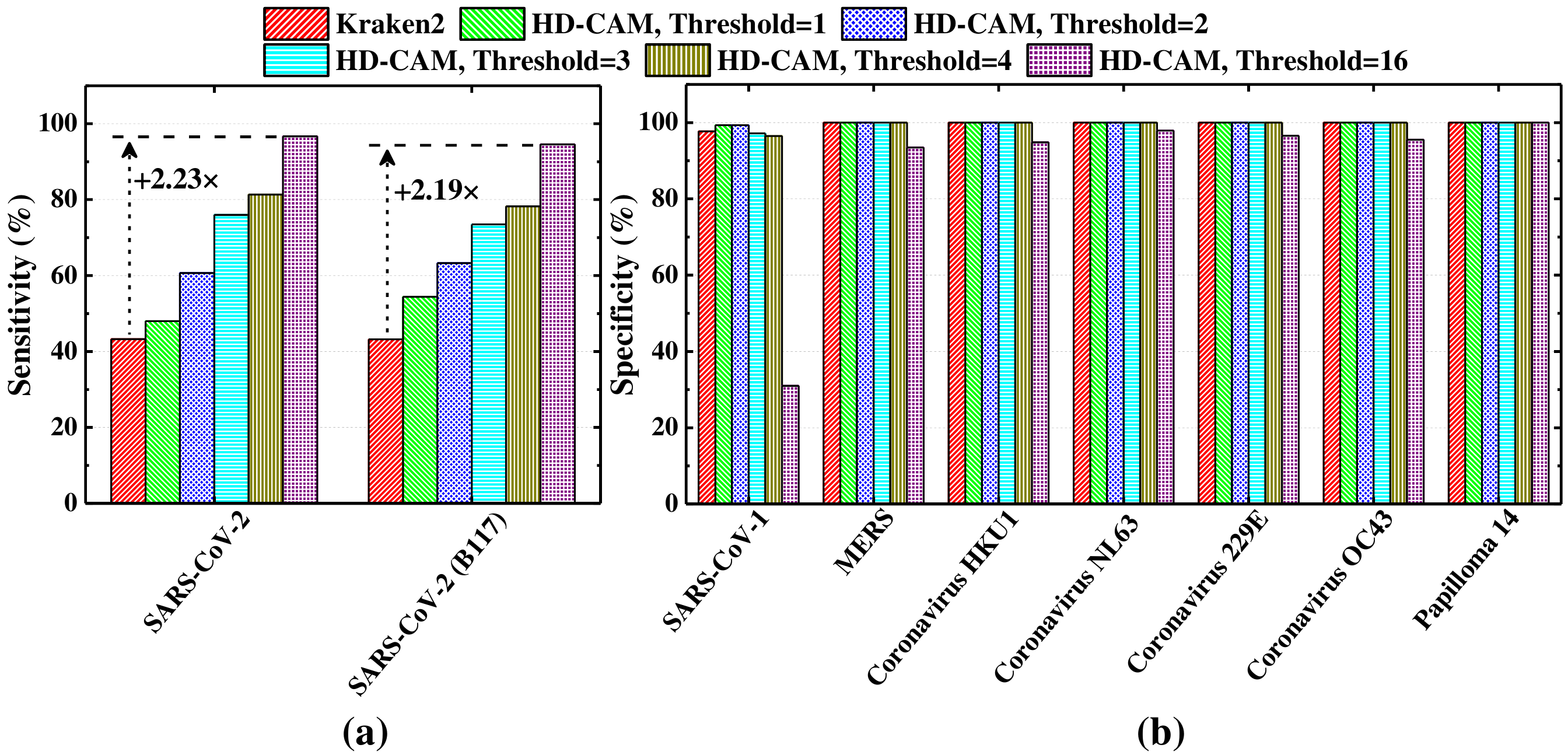}
	\caption{\black{(a) Sensitivity and (b) specificity of Kraken2 and HD-CAM virus DNA classification (Threshold is mismatch threshold [in basepairs]); The second (green) bar reflects the potential result of state of the art tunable sampling time technique where the mismatch threshold is limited to four bits.}}
	\label{fig:Application Results}
\end{figure*}

A sequenced sample (output of the sequencer in \figref{fig: HD-CAM as DNA classifier}(a)) typically consists of a large set of DNA fragments (refer to \figref{fig: HD-CAM as DNA classifier}(a)), called DNA reads, sourced from DNAs of different organisms (e.g., bacteria and viruses) presented in the sample. To classify the virus DNA in the sequenced sample, each DNA read is searched in the HD-CAM (refer to query pattern in \figref{fig: HD-CAM as DNA classifier}(b)).
In other words, a DNA read is a query pattern and is compared against the entire reference DNA database simultaneously.
Thus, ideally, reads that belong to the target virus DNA, should match exactly in the HD-CAM.
However, DNA reads contain sequencing errors~\cite{khatamifard2017read} of the following three types: replacement or substitution (where a certain basepair is called incorrectly, i.e., replaced by a wrong one), insertion (where a basepair is inserted into existing DNA sequence), and deletion (where a basepair is deleted from existing DNA sequence).
Another source of difference between the target DNA reads and the reference DNA are genetic variations, which may occur in mutations, such as UK, South African or Delta variants of \covid.
Such variations result in a nonzero Hamming distance between a read and a reference fragment (\kmer) that would otherwise match exactly.

The ability of HD-CAM to tolerate large numbers of mismatching bits enables DNA classification, even when the target DNA reads have multiple sequencing errors or genetic variations. 
Moreover, by allowing the programmable mismatch threshold, HD-CAM supports a wide variety of sequencing error profiles. 

The reads sourced from DNAs of other organisms presented in the sample, are expected to exhibit significant difference vs. the target virus DNA, such that the Hamming distance between such reads and the reference \kmers should typically be higher than the mismatch threshold, which in turn allows classifying those reads as not \covid.

\subsection{\blue{Experimental methodology}}
\label{subsec:classifier method}
The target virus is \covid, downloaded from \ncbi online data sets~\cite{ncbi2021}.
We encode the basepairs using one-hot encoding (A=0001, C=0010, G=0100, T=1000), as presented in \figref{fig: HD-CAM as DNA classifier}(b).
By using this coding scheme, we ensure that any basepair difference results in a \hd of two bits, regardless of the basepair value (A, T, G or C). 
Another alternative is to apply Gray coding. 
By selecting 3-bit Gray code values that are two positions apart (e.g., A=000, C=011, G=110 and T=101), we ensure that the \hd between any two basepair values is a constant two bits.  
For a 256-bit wide HD-CAM, the \kmer length is set at $k=64$.
Accordingly, the size of the reference DNA database in the HD-CAM is 29,903\Xnospace 256 bits.

We test three types of DNA samples.
The first type is created by extracting reads from random positions in the reference \covid DNA and injecting random errors into those reads
as follows: replacement rate = 3.6\%, insertion rate = 0.2\%, deletion rate = 0.2\%~\cite{rhoads2015pacbio}. 
\blue{It should be noted that while the effect of replacement rate on the Hamming distance is straightforward, 
even a single insertion or deletion may create a very large Hamming distance between the query \kmer and the reference \kmers.}   

The second type of sample is formed by extracting reads from the DNA of \covid UK variant.

Finally, the samples of the third type are created using DNA reads from other human coronaviruses (SARS, MERS, alpha, and beta), and Papilloma, which is not a coronavirus.

\blue{We expect the DNA reads of the first two types to be classified as \covid, while the reads of the third type should be classified as not \covid.}



\subsection{\blue{DNA classification evaluation}}
\label{subsec:DNA eval}
\blue{Similarly to the evaluation methodology introduced in \secref{sec:Evaluation}, we use sensitivity and specificity to evaluate the efficiency of virus classification by HD-CAM. Note that from this point and throughout the rest of the paper, sensitivity and specificity measure the probability of correct classification of a DNA read (as \covid or not \covid) rather than correct match or mismatch in general.} 

\blue{Here, $sensitivity=TP/(TP+FN)$, where  
$TP$ (true positive) means that target virus DNA read is correctly classified as \covid, and $FN$ (false negative) means that target virus DNA read is wrongly rejected. We use sensitivity to evaluate the efficiency of HD-CAM in analyzing the DNA of the first and the second type, because in these tests, no negative results are expected, hence all positive results are true and all negative results are false.} 

\blue{$Specificity=TN/(TN+FP)$, where  
$TN$ (true negative) means that not \covid read is correctly rejected, and $FP$ (false positive) means that unrelated DNA read is wrongly classified as \covid. We use specificity to evaluate the efficiency of HD-CAM in processing the DNA of the third type, because in such a test, no positive results are expected, hence all negative results are true and all positive results are false.}



We tested several DNA samples using different HD-CAM mismatch threshold values, i.e., the \hd (in basepairs) that the HD-CAM is configured to tolerate.

\blue{The resulting HD-CAM sensitivity (of \covid and its UK variant classification) and specificity (while processing DNA of other organisms) are presented in \figref{fig:Application Results}. For comparison, we also show the sensitivity and specificity of state of the art DNA classification tool Kraken2~\cite{wood2019improved}. }

\blue{We make the following observations:}
\begin{itemize}
    \item \blue{HD-CAM DNA classification sensitivity grows with the mismatch threshold, reaching 98\% when mismatch threshold is set at 16 basepairs, providing $2.2\times$ improvement vs. Kraken2;}

\item \blue{Kraken2 provides the lowest DNA classification sensitivity, because it employs exact rather than approximate search;}

\item \blue{The 4 bits~\cite{imani2017exploring} limitation on mismatch threshold results in limited DNA classification sensitivity (green bar in ~\figref{fig:Application Results});}

\item \blue{Insertions and deletions significantly reduce the sensitivity of Kraken2 as well as state of the art tunable sampling time schemes~\cite{Rahimi2015Approximate,imani2017exploring}, and prevent HD-CAM from reaching 100\% sensitivity;}

\item \blue{HD-CAM and Kraken2 exhibit similar specificity of nearly 100\% except for the case of SARS-CoV-1 (a virus very similar to \covid) and mismatch threshold of 16 basepairs, where the specificity drops due to the increased sensitivity.}

\end{itemize}
\section{Conclusion and Future Work}
\label{sec:conclusions}
In this paper we present a novel content addressable memory, HD-CAM, which enables a single cycle approximate search with programmable mismatch threshold (tolerable Hamming distance between the query and the stored pattern).
HD-CAM makes use of NOR-type bitcells that have been modified to allow approximate search by matchline charge redistribution.
Our design was implemented and evaluated with a 65\nm commercial CMOS technology \black{at different process corners and under process variation conditions} through extensive Monte Carlo simulations.
\black{Results demonstrate that HD-CAM exhibits very high sensitivity and specificity in a very wide range of dynamically-configurable mismatch threshold levels.}
\blue{We perform a detailed design space exploration and show that HD-CAM has very low susceptibility to process and sampling time variation. We also show that HD-CAM enables a wide range of pattern lengths and mismatch threshold levels, unsupported by state of the art approximate search CAM designs.}

\blue{HD-CAM is applied to DNA classification, which is one of the critical steps in genomic surveillance, employed throughout the world to combat Covid-19. Specifically, we use HD-CAM to classify \covid in a large genomic samples. HD-CAM is shown to outperform state of the art DNA classification tool in terms of sensitivity and specificity.} 

Finally, we would like to point out another potential application for HD-CAM: an ECC enabled CAM. Typically, using ECC in CAM is not trivial; if even a single bit of a memory block that stores an ECC-encoded search pattern fails, the search pattern would not match and a query would incorrectly result in a mismatch.  
HD-CAM potentially makes ECC protection easy. \blue{If the memory content is encoded such that the Hamming distance between any two valid codewords is larger than the worst case uncertainty region (where false matches and false mismatches are possible), then any error where the number of failed bits is below such uncertainty period, can be tolerated with 100\% guarantee. Hence, the number of tolerable errors (failed bits) in a single memory block can be varied by programming the mismatch threshold.} This application of our proposal will be evaluated in the future. 




\bibliographystyle{IEEEtran}
\bibliography{Bibliography}


\end{document}